# Interacting topological magnons in the Kitaev-Heisenberg honeycomb ferromagnets with Dzyaloshinskii-Moriya interaction

Jie Wang, Pei Chen, Bing Tang*

*Department of Physics, Jishou University, Jishou 416000, China*

ABSTRACT

This theoretical work is devoted to investigating the magnon-magnon interaction effect in a two-dimensional Heisenberg-Kitaev honeycomb ferromagnet with Dzyaloshinskii-Moriya interaction (DMI). Based on the first-order Green's function formalism, we calculate the thermal-fluctuation-induced temperature-dependent self-energy corrections of magnons. Our calculations reveal that the critical temperature for temperature-induced topological phase transitions monotonically approaches the Curie temperature with increasing DMI strength. Furthermore, it is shown that the critical temperature for topological phase transitions is correlated with Dzyaloshinskii-Moriya interaction and magnetic field strength.

## 1. Introduction

In the last 20 years, many researchers have focused on topological materials, such as topological insulators and semimetals [1-6]. The concept of topological phases has expanded from electronic systems to those bosonic systems, e.g., photonic, and magnon systems [2-6]. In certain magnetic materials, electrically neutral magnons can exhibit edge or surface states, which are resilient to disorder and backscattering. Theoretically, they can achieve long-range transmission without energy loss. Relevant research results do not only deepened our understanding of topological phases but also provide new insights and methodologies for designing electronic and magnon devices with low dissipation and low power consumption [5-6]. Such devices hold significant potential for future low-energy computing and communication technologies.

* Corresponding author.
E-mail addresses: bingtangphy@jsu.edu.cn

Recently, two-dimensional Heisenberg-Kitaev (HK) honeycomb ferromagnets have received much attention due to their abundant topological characteristics [7-16]. It has been proposed that the Kitaev interaction might co-occur with the DMI in some real magnetic materials [8]. For example, R. Jaeschke *et al.* [9] have conducted a theoretical investigation into the microscopic origins of anisotropic ferromagnetism in a van der Waals magnet ($CrI_3$). Their results have showed that those first nearest-neighbor exchange interactions can be effectively described by the HK-Γ model, and a non-zero DMI has been identified between next-nearest neighbors. What is more, they have found that the DMI plays a dominant role in determining the band gap at the Dirac point even though it is much weaker than the Kitaev interaction. In fact, by examining the transport properties of magnons, one may determine whether the magnetic system is primarily influenced by the DMI or the Kitaev interaction [15-16]. By using $CrI_3$ as a prototype to simulate two-dimensional van der Waals magnetic materials, Verena Brehm *et al*. have investigated the chiral magnon edge states generated by DM and Kitaev spin interactions [17].Their results show that these edge states can maintain robust. Although many theoretical works on the topological properties of HK honeycomb ferromagnets have been reported, few researchers have considered magnon-magnon interactions. Up to now, some works have been devoted to studying the effects of magnon-magnon interactions in honeycomb Heisenberg ferromagnets [20-24]. It has been shown that the magnon-magnon interaction can cause significant momentum-dependent renormalization of the band structure [20]. Especially, there are some evidences that magnon-magnon interactions can cause topological phase transitions of Dirac magnons in Heisenberg honeycomb ferromagnet [21-23].

In this letter, we shall investigate the effects of magnon-magnon interactions and related topological phase transitions in a two-dimensional HK honeycomb ferromagnet with the DMI. Utilizing the Green's function method, the effects of magnon-magnon interactions on the topological and transport properties of magnons are systematically analyzed. Using a self-consistent treatment, the average magnetization, magnon bands and the thermal Hall coefficient are calculated, respectively. Our results show that DMI introduction is a prerequisite for topological phase transitions in the constructed model. For a specific value of D, the topological phase of the ferromagnetic system can be effectively regulated by adjusting temperature or external magnetic field. Furthermore, increasing DMI strength monotonically drives the critical temperature of temperature-driven topological phase transitions toward the Curie temperature. Additionally, our results present rich magnonic topological phase diagrams as functions of DMI, temperature, and magnetic field strength. More details on the present study will be fully shown in the following sections.

## 2. Model

Let us take into account an extended ferromagnetic HK model on a two-dimensional honeycomb (monolayer) lattice (see Fig. 1). After being subjected to a uniform magnetic field B, the corresponding Hamiltonian can be written as [16].

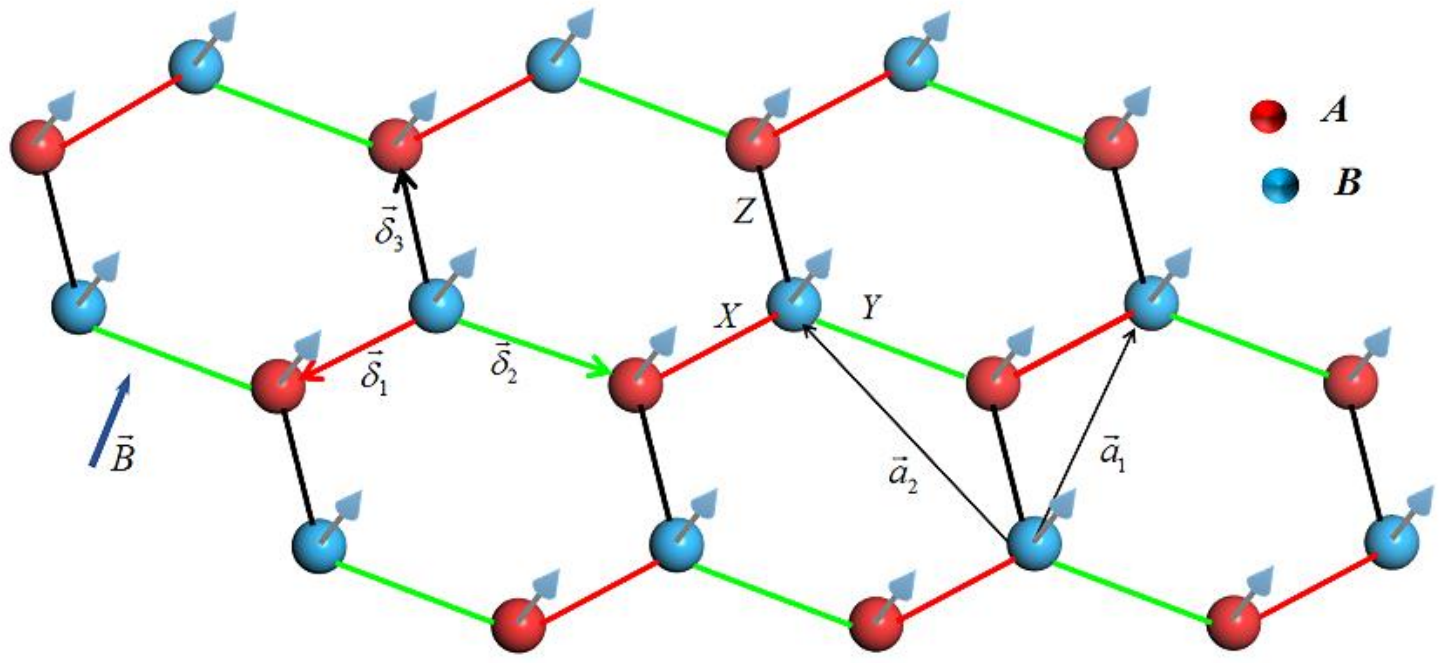

**Fig. 1.** Schematic illustration of an extended ferromagnetic HK model on the honeycomb ferromagnet lattice. The colors on the bonds represent the Kitaev bonds X (red), Y (green), Z ( black).

$$\mathcal{H} = J_H \sum_{\langle i,j\rangle} \mathbf{S}_i \cdot \mathbf{S}_j + 2J_K \sum_{\langle i,j\rangle\gamma} S_i^{\gamma} S_j^{\gamma} + \sum_{\langle\langle ij\rangle\rangle} \mathbf{D}_{ij} \cdot \left(\mathbf{S}_i \times \mathbf{S}_j\right) - \sum_i K_Z \left(\mathbf{e_z} \cdot \mathbf{S}_i\right)^2 - \mu_B g \mathbf{B} \cdot \sum_i \mathbf{S}_i \,. \quad (1)$$

The first two terms in Eq. (1) stand for the first-neighbor Heisenberg exchange and bond-dependent Kitaev interactions, respectively. The index $\gamma \in \{X,Y,Z\}$ represents the different links, as depicted in Fig. 1. Here, we have adopted the parametrization $J_H = A\cos\theta < 0$, and $J_K = A\sin\theta$, and $A > 0$ corresponds to one overall energy scale [25] . The third term represents the second nearest neighbors DMI. $\mathbf{D}_{ij} = D v_{ij} \mathbf{e_z}$ is known as the DMI vector, where $v_{ij} = -v_{ji} = \pm 1$ is contingent upon the orientation of spins in the two second-nearest neighbors and $\mathbf{e_z}$ is applied parallel to [111] direction. Additionally, identified as the easy axis. The last term stands for the coupling to the magnetic field $\mathbf{B} = B\mathbf{e_z}$. For convenience, one can define $h \equiv \mu_B \mathrm{g} \mathbf{B}$.

## 3. Interacting topological Dirac magnons

### *3.1. Methodology.*

Many theoretical works on topological magnons have been confined to the linear spin wave theory. At low temperatures, this treatment is reasonable. But in actuality, with the increase of the temperature, the role of magnon-magnon interactions become more and more important so that the higher-order effect in the renormalized spin wave theory cannot be ignored. In order to obtain the quantized form of the HK model Hamiltonian (1), we make use of the Holstein-Primakoff transformation to rewrite the spin operators [26]. Subsequently, by performing a Fourier transformation, we can obtain the Hamiltonian including a non-interacting term $\mathcal{H}^{(2)}$ and an interacting term $\mathcal{H}_{\text{int}}$. The non-interacting Hamiltonian $\mathcal{H}^{(2)}$ can be written in the form $\mathcal{H}^{(2)}=\frac{1}{2}\sum_{\mathbf{k}}\psi_{\mathbf{k}}^{+}H_0(\mathbf{k})\psi_{\mathbf{k}}$, where $\psi_{\mathbf{k}}=\left(a_{\mathbf{k}},b_{\mathbf{k}},a_{-\mathbf{k}}^{+},b_{-\mathbf{k}}^{+}\right)^T$ Here, the Hamiltonian matrix $H_0(\mathbf{k})$ is given by

$$H_0(\mathbf{k})=\begin{pmatrix} A_{\mathbf{k}}^0 & B_{\mathbf{k}}^0 \\ \left[B_{\mathbf{k}}^0\right]^{\dagger} & \left[A_{-\mathbf{k}}^0\right]^T \end{pmatrix}, \tag{2}$$

with

$$A_{\mathbf{k}}^0=\begin{pmatrix} \rho_{a,\mathbf{k}} & \rho_{1,\mathbf{k}} \\ \rho_{1,\mathbf{k}}^* & \rho_{b,\mathbf{k}} \end{pmatrix} \ , \ B_{\mathbf{k}}^0=\begin{pmatrix} 0 & \rho_{2,\mathbf{k}} \\ \rho_{2,-\mathbf{k}} & 0 \end{pmatrix} \ , \tag{3}$$

where

$\rho_{a,\mathbf{k}}/\rho_{b,\mathbf{k}}=-3SJ_0+K_Z\left(2S-1\right)+h\pm\frac{2S}{3}Dd_{\mathbf{k}}$ , $\rho_{1,\mathbf{k}}=J_0f_{\mathbf{k}}^*$ , $\rho_{2,\mathbf{k}}=\frac{2}{3}J_K g_{\mathbf{k}}^*$ , $J_0=J_H+2J_K/3$ , $f_{\mathbf{k}}=(1+e^{i\mathbf{k}\cdot\boldsymbol{a}_1}+e^{i\mathbf{k}\cdot\boldsymbol{a}_2})$ , $g_{\mathbf{k}}=(1+e^{i(\mathbf{k}\cdot\boldsymbol{a}_1+2\pi/3)}+e^{i(\mathbf{k}\cdot\boldsymbol{a}_2-2\pi/3)})$ , and $d_{\mathbf{k}}=\sum_{n=1}^{3}\sin(\mathbf{k}\cdot\varsigma_n)$.

The interacting Hamiltonian $\mathcal{H}_{\text{int}}$ reads

$$
\begin{aligned}
\mathcal{H}_{\text{int}} = \frac{1}{N}\sum_{\{\mathbf{k}_i\}} \Big\{ & -\frac{J_0}{4}\Big( f^{*}_{\mathbf{k}_4} a^{+}_{\mathbf{k}_1} a^{+}_{\mathbf{k}_2} a_{\mathbf{k}_3} b_{\mathbf{k}_4} \\
& + f^{*}_{\mathbf{k}_4} b_{\mathbf{k}_1} b^{+}_{\mathbf{k}_2} b^{+}_{\mathbf{k}_3} a^{+}_{\mathbf{k}_4} - 4 f^{*}_{\mathbf{k}_4-\mathbf{k}_2} a^{+}_{\mathbf{k}_1} b^{+}_{\mathbf{k}_2} a_{\mathbf{k}_3} b_{\mathbf{k}_4} + H.c \Big) \delta^{1}_{\{\mathbf{k}_i\}} \\
& - \frac{J_K}{6}\Big( g^{*}_{\mathbf{k}_4} b_{\mathbf{k}_1} b^{+}_{\mathbf{k}_2} b^{+}_{\mathbf{k}_3} a^{+}_{\mathbf{k}_4} + g^{*}_{-\mathbf{k}_4} a_{\mathbf{k}_1} a^{+}_{\mathbf{k}_2} a^{+}_{\mathbf{k}_3} b^{+}_{\mathbf{k}_4} + H.c \Big) \delta^{2}_{\{\mathbf{k}_i\}} \\
& + \Big[ \big( m_{\mathbf{k}} - K_Z \big) a^{+}_{\mathbf{k}_1} a^{+}_{\mathbf{k}_2} a_{\mathbf{k}_3} a_{\mathbf{k}_4} + \big( -m_{\mathbf{k}} - K_Z \big) b^{+}_{\mathbf{k}_1} b^{+}_{\mathbf{k}_2} b_{\mathbf{k}_3} b_{\mathbf{k}_4} \Big] \delta^{1}_{\{\mathbf{k}_i\}} \Big\},
\end{aligned} \tag{4}
$$

where $\delta^{1}_{\{\mathbf{k}_i\}} = \delta_{\mathbf{k}_1+\mathbf{k}_2,\mathbf{k}_3+\mathbf{k}_4}$, $\delta^{2}_{\{\mathbf{k}_i\}} = \delta_{\mathbf{k}_1,\mathbf{k}_2+\mathbf{k}_3+\mathbf{k}_4}$, and $m_{\mathbf{k}} = \frac{D}{6}(d_{\mathbf{k}_2} + d_{\mathbf{k}_4})$. In order to take into account the effect of magnon-magnon interactions at finite temperature, we apply the Green's function method [41]. Firstly, a matrix Green's function can be defined via $\mathcal{G}(\mathbf{k},\tau) = -\left\langle \mathcal{T}_\tau \psi_{\mathbf{k}}(\tau) \psi^{+}_{\mathbf{k}}(0) \right\rangle$. Here, the angle bracket $\langle \cdots \rangle$ stands for the thermodynamic average, and $\mathcal{T}_\tau$ corresponds to a chronological operator with respect to the imaginary time $\tau = it$ ( $0 \le \tau \le \beta$ ), where $\beta = \left(k_B T\right)^{-1}$. In the Heisenberg picture, a time-dependent operator can be express as $O\left(\tau\right) = e^{\mathcal{H}\tau} O(0) e^{-\mathcal{H}\tau}$, where $\mathcal{H} = \mathcal{H}^{(2)} + \mathcal{H}_{\text{int}}$ is the Hamiltonian of the system. In order to obtain an effective quadratic Hamiltonian, one can solve the Heisenberg equation of motion of the Green's function and make use of the random phase approximation so as to get the nonlinear self-energy correction resulting from the magnon-magnon interactions. In principle, the Heisenberg equation of motion of the Green's function can be written in the following form

$$
\frac{d\mathcal{G}(\mathbf{k},\tau)}{d\tau} = -\delta(\tau)\tau_z - \tau_z\left( H_0(\mathbf{k}) + \sum\nolimits_{\mathbf{k}} \right) \mathcal{G}(\mathbf{k},\tau). \tag{5}
$$

Here, $\tau_z$ represents a Pauli matrix acting on particle-hole space [17], and $\sum_{\mathbf{k}}$ is known as the self-energy term, which is derived from the random phase approximation. Subsequently, we perform the Fourier transformation $\mathcal{G}(\mathbf{k},\tau) = \frac{1}{\beta}\sum_{n} e^{-i\omega_n \tau} \mathcal{G}(\mathbf{k},\omega_n) e^{-i\omega_n \tau}$, where $\omega_n$ stands for the socalled bosonic

Matsubara frequency. Through some simple derivations, one can obtain the equation $-i\omega_n \mathcal{G}(\mathbf{k},\omega_n) = -\tau_z - \tau_z \left(H_0(\mathbf{k}) + \sum_{\mathbf{k}}\right)\mathcal{G}(\mathbf{k},\omega_n)$. Multiplying both sides of the equation by the $\tau_z$ term, then one can get the Dyson's equation $\mathcal{G}^{-1}(\mathbf{k},\omega_n) = i\omega_n \tau_z - H_{\mathbf{k}}^{eff}$, where the renormalized effective Hamiltonian $H_{\mathbf{k}}^{eff}$ is given by

$$H_{\mathbf{k}}^{eff} = \begin{pmatrix} A_{\mathbf{k}}^{eff} & B_{\mathbf{k}}^{eff} \\ \left[B_{\mathbf{k}}^{eff}\right]^{\dagger} & \left[A_{-\mathbf{k}}^{eff}\right]^{T} \end{pmatrix}, \tag{6}$$

with

$$A_{\mathbf{k}}^{eff} = \begin{pmatrix} \rho_{a,\mathbf{k}}^{eff} & \rho_{1,\mathbf{k}}^{eff} \\ (\rho_{1,\mathbf{k}}^{eff})^{*} & \rho_{b,\mathbf{k}}^{eff} \end{pmatrix}, \qquad B_{\mathbf{k}}^{eff} = \begin{pmatrix} 0 & \rho_{2,\mathbf{k}}^{eff} \\ \rho_{2,-\mathbf{k}}^{eff} & 0 \end{pmatrix}, \tag{7}$$

where

$$\rho_{a,\mathbf{k}}^{eff} = -3J_0\bar{S}_B + \frac{2D}{3}\bar{S}_A d_{\mathbf{k}} + h_{a,b} + K_Z\left(-2S - 1 + 4\bar{S}_A\right), \tag{8}$$

$$\rho_{b,\mathbf{k}}^{eff} = -3J_0\bar{S}_A - \frac{2D}{3}\bar{S}_B d_{\mathbf{k}} + h_{a,b} + K_Z\left(-2S - 1 + 4\bar{S}_B\right), \tag{9}$$

$$\rho_{1,\mathbf{k}}^{eff} = \frac{J_0}{2}\left(\bar{S}_A + \bar{S}_B\right) f_{\mathbf{k}}^{*} + \frac{J_0}{N}\sum_{q} f_{\mathbf{k}-\mathbf{q}}^{*}\left\langle b_{\mathbf{q}}^{+} a_{\mathbf{q}}\right\rangle, \tag{10}$$

$$\rho_{2,\mathbf{k}}^{eff} = \frac{1}{3}J_K\left(\bar{S}_A + \bar{S}_B\right) g_{\mathbf{k}}^{*} + J_0\frac{1}{N}\sum_{q} f_{\mathbf{k}-\mathbf{q}}^{*}\left\langle a_{\mathbf{q}} b_{-\mathbf{q}}\right\rangle, \tag{11}$$

$$\bar{S}_A = S - \frac{1}{N}\sum_{\mathbf{q}}\left\langle a_{\mathbf{q}}^{+} a_{\mathbf{q}}\right\rangle, \quad \bar{S}_B = S - \frac{1}{N}\sum_{\mathbf{q}}\left\langle b_{\mathbf{q}}^{+} b_{\mathbf{q}}\right\rangle, \tag{12}$$

$$h_{a,b} = -\frac{J_0}{N}\sum_{\mathbf{q}}\mathrm{Re}\left(f_{\mathbf{q}}^{*}\left\langle a_{\mathbf{q}}^{+} b_{\mathbf{q}}\right\rangle\right) - \frac{2J_K}{3N}\sum_{\mathbf{q}}\mathrm{Re}\left(g_{\mathbf{q}}^{*}\left\langle a_{\mathbf{q}}^{+} b_{-\mathbf{q}}^{+}\right\rangle\right) + h, \tag{13}$$

Within the random phase approximation, the effect of higher-order terms on the physical properties of the system is very weak and are thus omitted [42]. By means of diagonalizing the effective Hamiltonian (6), i.e., $\Lambda_{\mathbf{k}}^{\dagger} H_{\mathbf{k}}^{eff} \Lambda_{\mathbf{k}} = diag\{E_{\mathbf{k}}, E_{-\mathbf{k}}\}$, one can obtain $H_{\mathbf{k}}^{eff} = \sum_{\mathbf{k}}\left(E_{\mathbf{k}}^{\alpha}\alpha_{\mathbf{k}}^{\dagger}\alpha_{\mathbf{k}} + E_{\mathbf{k}}^{\beta}\beta_{\mathbf{k}}^{\dagger}\beta_{\mathbf{k}}\right)$ up to the zero-point energy. Here, the paraunitary matrix need to obey $\Lambda_{\mathbf{k}}^{\dagger}\tau_z\Lambda_{\mathbf{k}} = \tau_z$. It is noted that the diagonalization process yields a relation $\psi_{\mathbf{k}} = \Lambda_{\mathbf{k}} \phi_{\mathbf{k}}$, where $\phi_{\mathbf{k}} = (\alpha_{\mathbf{k}}, \beta_{\mathbf{k}}, \alpha^{+}_{-\mathbf{k}}, \beta^{+}_{-\mathbf{k}})^{T}$. According to the paraunitary matrix $\Lambda_{\mathbf{k}}$, one can get the following relational expressions

$$a_{\mathbf{k}} = u_{\mathbf{k},a,\alpha} \alpha_{\mathbf{k}} + u_{\mathbf{k},a,\beta} \beta_{\mathbf{k}} + v_{-\mathbf{k},a,\alpha} \alpha^{+}_{-\mathbf{k}} + v_{-\mathbf{k},a,\beta} \beta^{+}_{-\mathbf{k}}, \tag{14}$$

$$b_{\mathbf{k}} = u_{\mathbf{k},b,\alpha} \alpha_{\mathbf{k}} + u_{\mathbf{k},b,\beta} \beta_{\mathbf{k}} + v_{-\mathbf{k},b,\alpha} \alpha^{+}_{-\mathbf{k}} + v_{-\mathbf{k},b,\beta} \beta^{+}_{-\mathbf{k}}. \tag{15}$$

Via a Bogoliubov transformation, we can obtain the expectation values corresponding to the pair correlators, which can be expressed in the following forms

$$\begin{aligned}
\sum_{\mathbf{q}} \left\langle a^{+}_{\mathbf{q}} a_{\mathbf{q}} \right\rangle &= \frac{1}{N} \sum_{\mathbf{q}} \left( \left| u_{\mathbf{q},a,\alpha} \right|^2 \left\langle \alpha^{+}_{\mathbf{q}} \alpha_{\mathbf{q}} \right\rangle + \left| u_{\mathbf{q},a,\beta} \right|^2 \left\langle \beta^{+}_{\mathbf{q}} \beta_{\mathbf{q}} \right\rangle + \left| v_{-\mathbf{q},a,\alpha} \right|^2 \left\langle \alpha_{-\mathbf{q}} \alpha^{+}_{-\mathbf{q}} \right\rangle + \left| v_{-\mathbf{q},a,\beta} \right|^2 \left\langle \beta_{-\mathbf{q}} \beta^{+}_{-\mathbf{q}} \right\rangle \right) \\
&= \frac{1}{N} \sum_{\mathbf{q}} \left( \left| u_{\mathbf{q},a,\alpha} \right|^2 n_{q,\alpha} + \left| u_{\mathbf{q},a,\beta} \right|^2 n_{\mathbf{q},\beta} + \left| v_{-\mathbf{q},a,\alpha} \right|^2 \left( 1 + n_{-\mathbf{q},\alpha} \right) + \left| v_{-\mathbf{q},a,\beta} \right|^2 \left( 1 + n_{-\mathbf{q},\beta} \right) \right) \\
&= \sum_{\mathbf{q},\lambda=\alpha,\beta} \left( \left| u_{\mathbf{q},a,\lambda} \right|^2 + \left| v_{\mathbf{q},a,\lambda} \right|^2 \right) n_{\mathbf{q},\lambda} + \left| v_{\mathbf{q},a,\lambda} \right|^2, \\
\sum_{\mathbf{q}} \left\langle b^{+}_{\mathbf{q}} b_{\mathbf{q}} \right\rangle &= \sum_{\mathbf{q},\lambda=\alpha,\beta} \left( \left| u_{\mathbf{q},b,\lambda} \right|^2 + \left| v_{\mathbf{q},b,\lambda} \right|^2 \right) n_{\mathbf{q},\lambda} + \left| v_{\mathbf{q},b,\lambda} \right|^2, \\
\sum_{\mathbf{q}} \left\langle a^{+}_{\mathbf{q}} b^{+}_{-\mathbf{q}} \right\rangle &= \sum_{\mathbf{q},\lambda=\alpha,\beta} \left( u^{*}_{\mathbf{q},a,\lambda} v^{*}_{\mathbf{q},b,\lambda} + v^{*}_{-\mathbf{q},a,\lambda} u^{*}_{-\mathbf{q},b,\lambda} \right) n_{\mathbf{q},\lambda} + v^{*}_{-\mathbf{q},b,\lambda} u^{*}_{-\mathbf{q},b,\lambda}, \\
\sum_{\mathbf{q}} \left\langle a^{+}_{q} b_{q} \right\rangle &= \sum_{\mathbf{q},\lambda=\alpha,\beta} \left( u^{*}_{\mathbf{q},a,\lambda} u_{\mathbf{q},b,\lambda} + v^{*}_{\mathbf{q},a,\lambda} v_{\mathbf{q},b,\lambda} \right) n_{\mathbf{q},\lambda} + v^{*}_{\mathbf{q},a,\lambda} v_{\mathbf{q},b,\lambda},
\end{aligned} \tag{16}$$

where $n_{\mathbf{q},\alpha} = \left\langle \alpha^{+}_{\mathbf{q}} \alpha_{\mathbf{q}} \right\rangle = (e^{E^{\alpha}_{\mathbf{q}}/T} - 1)^{-1}$ and $n_{\mathbf{q},\beta} = \left\langle \beta^{+}_{\mathbf{q}} \beta_{\mathbf{q}} \right\rangle = (e^{E^{\beta}_{\mathbf{q}}/T} - 1)^{-1}$. Note that both $\left\langle \alpha^{+}_{\mathbf{q}} \alpha_{\mathbf{q}} \right\rangle$ and $\left\langle \beta^{+}_{\mathbf{q}} \beta_{\mathbf{q}} \right\rangle$ are the only non-zero thermal averages in the new foundation. Consistent with Refs.**错误!未找到引用源。**,**错误!未找到引用源。**, Eqs. (8) and (14), along with the mentioned diagonalization relationship. Using these equations, we can obtain the band structures and magnetization at specific temperatures in a self-consistent manner and calculate the corresponding Chern numbers.

*3.2. Magnetization.*

In ferromagnetic materials, the Curie temperature $T_c^*$ marks the threshold at which the material transitions from a ferromagnetic to a paramagnetic state. Here, due to the equivalence of the sublattices, we can introduce the quantity $M_z$ as an order parameter to character the magnetic phase transition. The average magnetization along the $\mathbf{e}_z$ direction is given by $M_z = \overline{S}$ [34]. Using a self-consistent calculation on the temperature dependence of the vanishing magnetization, one can directly determine the Curie temperature $T_c^*$ [33]. Fig. 2 shows the average magnetization as functions of the temperature for different magnetic fields. Under a constant magnetic field, the susceptibility tends to increase as the temperature decreases. Conversely, at a constant temperature, the susceptibility rises with an increase in the magnetic field. This phenomenon suggests that spin fluctuations are effectively suppressed by the appearance of a magnetic field. The Curie temperature increases as the magnetic field increases, since the Zeeman contribution results in variations of $T_c^*$ as one function of the magnetic field. These results demonstrate that the thermodynamic expectation value associated with the Bose-Einstein distribution exhibits a decreasing dependence on the strength of the applied magnetic field [40].

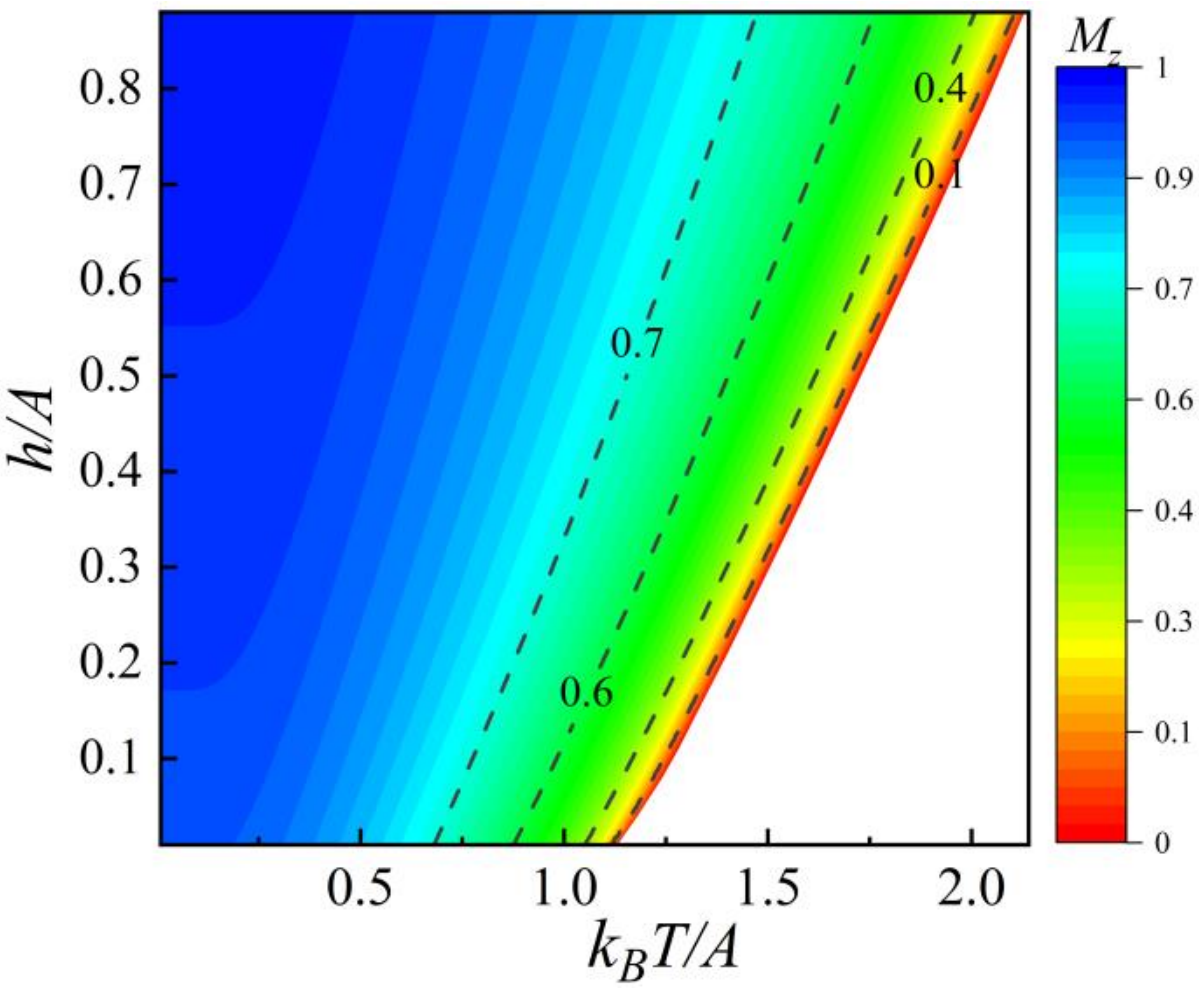

Fig. 2. The average magnetization $M_z$ distribution in the $k_BT-h$ plane under the [111] magnetic field. The remaining parameters are selected as $S=1$, $\theta$=1.45$\pi$, $K_Z$=0.15, and $D=0.3A$. The same parameters are utilized unless otherwise specified in subsequent calculations.

*3.3. Structures of renormalized magnon bands.*

As shown in Eq (16), the self-energy correction term has two components, a thermal one, proportional to the magnon number that vanishes at $T=0$, and a quantum one, that is finite at $T=0$错误!未找到引用源。. In the nonfrustrated ferromagnetic and antiferromagnetic Heisenberg cases the quantum contribution vanish, but they become enhanced by increasing the Kitaev exchange K 错误!未找到引用源。. The quantum contribution is a feature unique to Kitaev interactions.

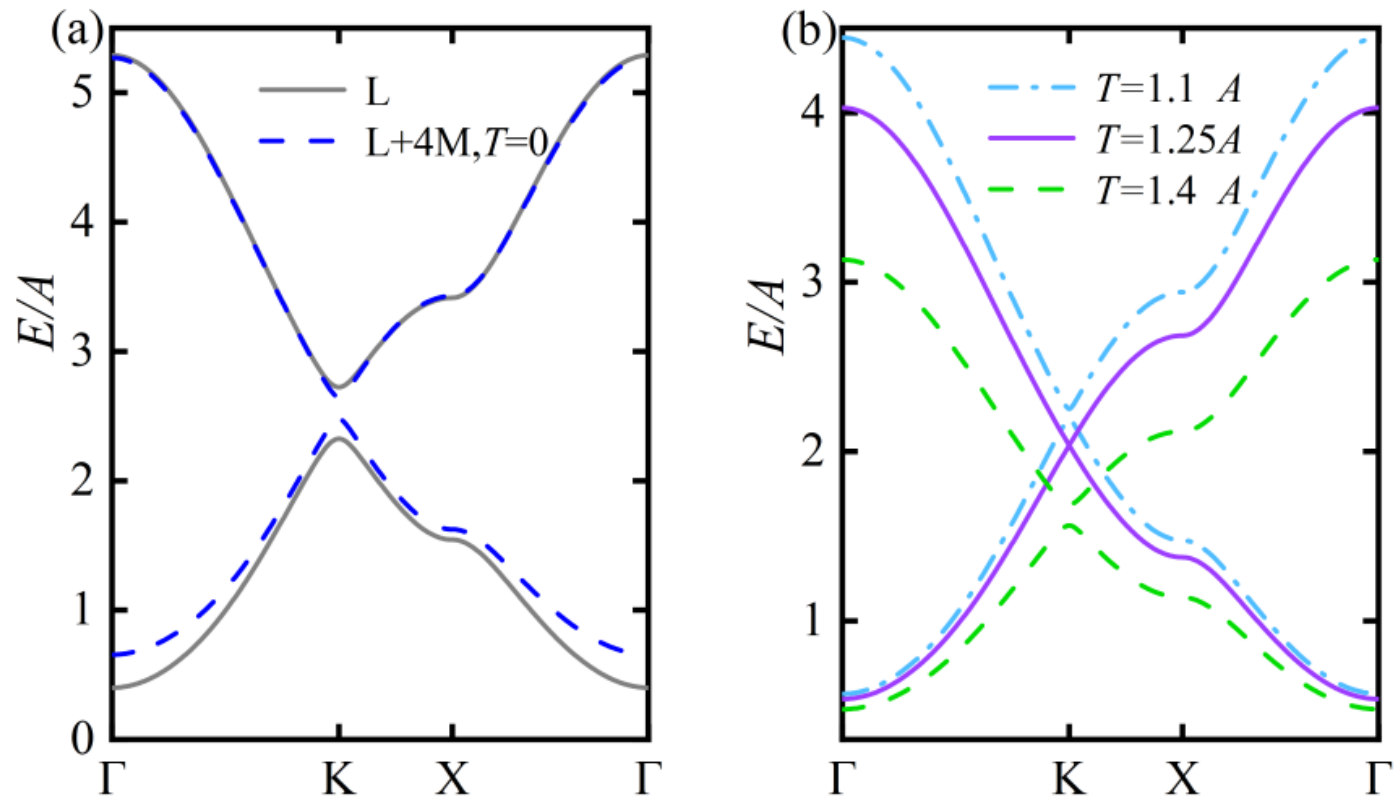

**Fig. 3.** Magnonic band structures along the high symmetry path $\Gamma-\mathrm{K}-\mathrm{X}-\Gamma$. (a) The noninteracting LSWT (L) approximation (solid gray lines) and in the presence of the four-magnon interactions (dashed blue lines) for $T=0$ and $h=0.25A$. (b) Magnonic band structures for three different temperatures.

In this work, we study the HK model at finite $T$ and low $h$. Fig. 3(a) illustrates the magnon bands at $T=0$. It can be observed that zero-point quantum fluctuations shift the acoustic branch of magnons upward relative to the noninteracting case (gray solid lines). Once thermal fluctuations appear, they decrease the effective magnetic exchange interaction, shifting the entire magnon band downward and thus narrowing the magnon bandwidth. The inclusion of magnon-magnon interactions is important, and renormalized spin-wave theory captures the essence of magnon softening or the temperature renormalization of the magnon band. It has been observed experimentally [46, 47], that the magnon excitation energy decreases as temperature increases, and the renormalization is especially strong when approaching $T_c$. Fig. 3(b) shows the temperature-dependent magnon bands under $h = 0.25A$ with a converged $T_C^* \approx 1.45A$ from renormalized spin-wave theory, presenting the calculated temperature-dependent magnon bands at $T = 1.1A, 1.25A, 1.4A$. As what can be seen in experiment, the calculated magnon excitation energy decreases as temperature increasing, and more rapidly at a higher temperature range. This is a direct consequence of spin-wave interactions where temperature changes the strength of the scattering. Renormalized spin-wave theory provides an excellent description of the temperature-induced renormalization effect, with its theoretical predictions enabling direct and quantitative comparison to the findings from neutron diffraction experiments [51]. These experimental techniques exhibit the capability to resolve band gaps down to the $\mu$eV energy scale.

Furthermore, we find that the renormalized magnon band gap at the K point exhibits thermal contraction and is entirely quenched at a well-defined temperature. Upon exceeding this temperature, the gap reopens and broadens with increasing temperature. From a topological perspective, the closure of the the renormalized magnon band gap is a necessary condition for the onset of a topological phase transition, while the band gap reopening marks the culmination of this phase transition process. Consequently, the subsequent priority will be to experimentally or theoretically confirm the existence of a topological phase transition within this temperature parameter regime.

*3.4. Phase diagrams and topological phase transitions.*

In principle, topological feature of magnon energy bands can be depict via the Chern number and thermal Hall conductivity, which are both represented as the discrete summation for the weighted Berry curvature over a series of points chosen properly stretching across the Brillouin zone. In topologically nontrivial phases, the Berry curvature displays marked spatial variations across momentum space. On the basis of the method proposed by Fukui *et al*. [42], one can efficiently calculate the Chern number for the λ-th band via

$$C_{\lambda}=\frac{1}{2\pi}\int_{BZ}d^{2}k\ B_{\lambda}^{z}(\mathbf{k}) \tag{17}$$

with the Berry curvature of the bands as $\mathbf{B}_{\lambda}(\mathbf{k})=\nabla\times\mathbf{A}_{\lambda}$ and the Berry connection $\mathbf{A}_{\lambda}=i\,\mathrm{Tr}[\Gamma^{\lambda}\Lambda_{\mathbf{k}}^{\dagger}\tau_{z}(\partial_{\mathbf{k}}\Lambda_{\mathbf{k}})]$, where $\lambda=+,-$ correspond to the top and the bottom of renormalized magnon bands, respectively. Upon calculating the Chern number associated with the energy bands. The gap reopens at $T=T_c$ suggests that there may be a topological phase transition at $T_c$. To confirm this, we compute the Chern number of each band above and below $T_c$ and verify that it indeed changes sign at $T_c$. Physically, magnon band closed at the critical value is fundamental to make sure the magnetic topological phase transition [4].

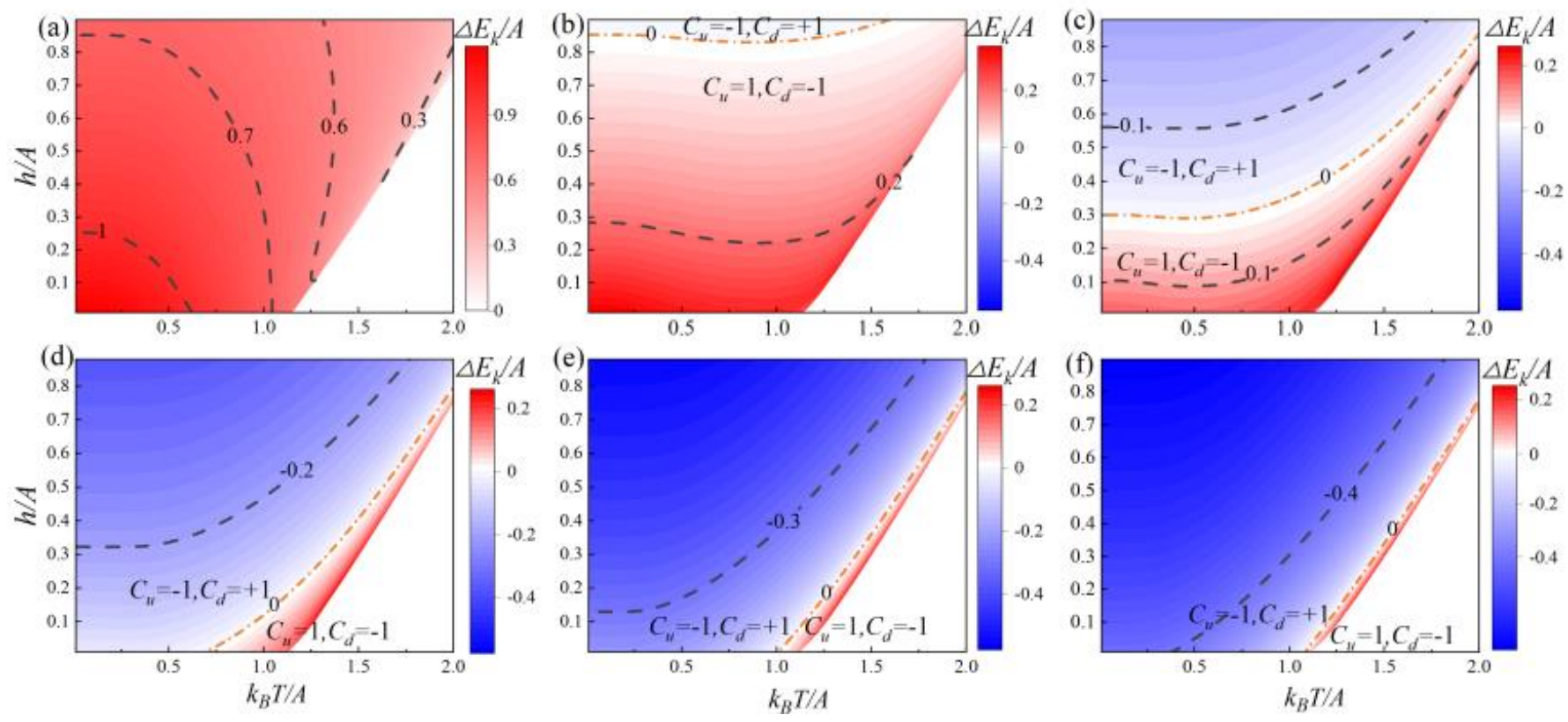

**Fig. 4.** The band gap $\Delta E_{\mathbf{K}}$ distribution in the $T-h$ plane for (a) $D=0.$, (b) $D=0.2A.$ (c) $D=0.25A.$ (d) $D=0.3A.$ (e) $D=0.35A.$ and (f) $D=0.4A.$ The orange dashed-dot line represents the condition where the band gap difference at the **K** point is zero.

In Fig. 4, we present the phase diagram of the renormalized magnon band gap as one function of the temperature and applied magnetic field for six different values of *D*. For definiteness, we plot the phase boundary between these two distinct topological phases where the Chern numbers in blue region (red region) are $C_u = -1(+1), C_d = +1(-1)$. In the $T-h$ phase diagram, the phase boundary is determined by the orange dashed-dot line, which is determined by examining the gap at the K point. This suggests that the topological phase of the present system can be effectively regulated by adjusting the temperature $T$ or the applied magnetic field strength $h$. These continuous topological phase transitions, which are driven by magnon-magnon interactions, are characterized by the occurrence of a gap-closing phenomenon. Moreover, further analysis revealed that as the magnetic field strength increases, critical temperature $T_c$ for band gap closure also rises. This phenomenon can be attributed to the stabilizing effect of the magnetic field on magnetic order, which enhances the stability of the ferromagnetic phase and thus requires a higher temperature to induce a phase transition.

Furthermore, we can observe that the DMI exerts a nontrivial modulation effect on the topological phase transition within the HK honeycomb ferromagnetic model. Specifically, for regimes with weak DMI strength, temperature-driven topological phase transitions are only accessible under moderately high magnetic field strengths. In stark contrast, for regimes with strong DMI strength, magnetic field-induced topological phase transitions demand elevated temperature environments to be realized. Theoretical analysis corroborates that the inclusion of DMI is an indispensable condition for the emergence of topological phase transitions in our model system. As depicted in Fig. 4(a), neither temperature variation nor magnetic field tuning can trigger topological phase transitions in this magnon system in the absence of DMI.

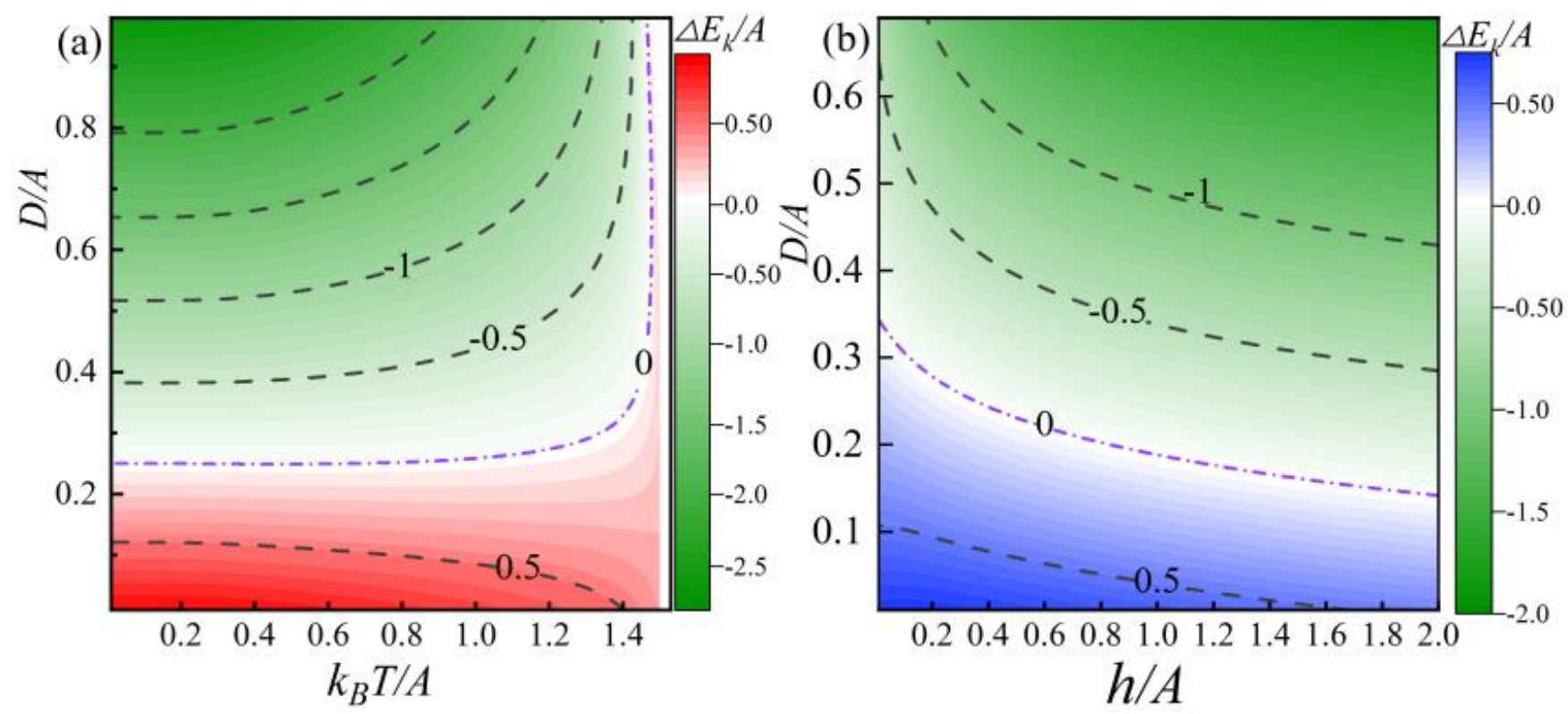

**Fig. 5.** (a). The band gap $\Delta E_{\mathbf{K}}$ as an order parameter of $T$ and $D$, with $h = 0.3A$. (b) The band gap $\Delta E_{\mathbf{K}}$ as an order parameter of $h$ and $D$, with $T = 1A$.

Below the Curie temperature, we proceed to examine the effects of magnetic field and temperature on the renormalized magnon band gap for different DMI strengths. In Fig. 5(a), we present the dependence of the energy gap on the temperature as well as DMI strength under relatively low magnetic field intensities.Our findings demonstrate that when the DMI strength exceeds a critical threshold (above the purple dash-dotted line), temperature tuning can trigger band gap closure and subsequent reopening, with the critical temperature for temperature-induced topological phase transitions monotonically approaching the Curie temperature with increasing DMI strength. In the elevated-temperature regime, we further map the evolution of the band gap with respect to the applied magnetic field strength and DMI strength (Fig. 5(b)), revealing that magnetic field strength can tailor the size of the energy gap at the K point. Additionally, for DMI strengths below the critical threshold (purple dash-dotted line), magnetic field tuning can drive reversible band gap closure and reopening. We confirm that all observed topological phase transitions occur at finite magnetization with , and these transitions retain their robustness, with their onset not being attributable to the system nearing a zero-magnetization state.

In two-dimensional magnetic systems, a longitudinal temperature gradient will cause one transverse heat flow, and the Berry curvature also acts as an effective magnetic field in the momentum space, giving rise to a non-quantized thermal magnon Hall effect. The associated thermal Hall conductivity can be written in the following form [38]

$$\kappa_{xy} = -\frac{k_B^2 T}{(2\pi)^2 \hbar} \sum_{\lambda=\alpha,\beta} \int d^2\mathbf{k} B_\lambda^z(\mathbf{k}) c_2(n_{\mathbf{k},\lambda}), \tag{16}$$

where $k_B^2$ is the Boltzmann constantis, $\hbar$ is the reduced Planck constant, $c_2(x) = (1+x)\left(\ln\frac{1+x}{x}\right)^2 - (\ln x)^2 - 2\mathrm{Li}_2(-x)$, where $\mathrm{Li}_2(-x)$ is the polylogarithm function. We will rigorously demonstrate in subsequent sections that the sign of the thermal Hall conductivity is identical to that of the Chern number corresponding to the up-spin magnon band.

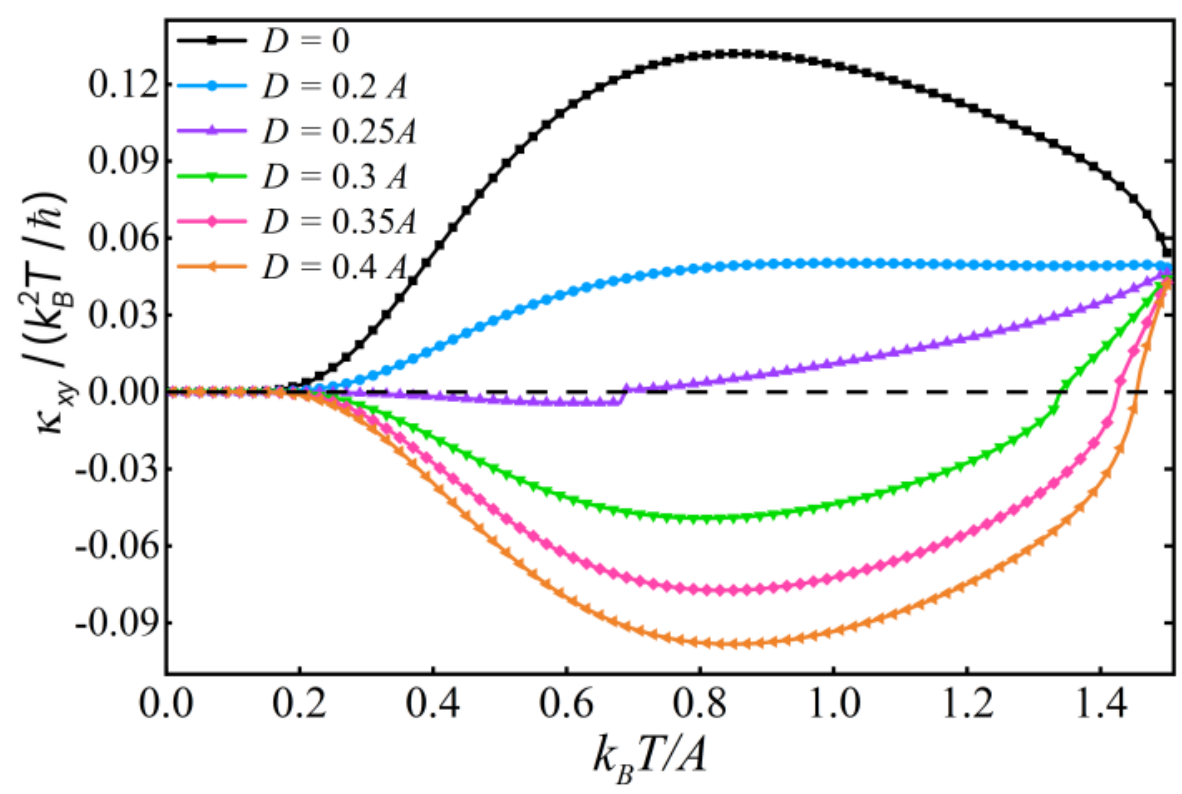

**Fig. 6.** The thermal Hall conductivity vs temperature for different values of $D$. The magnetic field is set to $h = 0.3A$.

Fig. 6 presents the temperature dependence of $\kappa_{xy}$ for different DMI strengths at a magnetic field of 0.3A. Consistent with the preceding discussion, the magnitude of $\kappa_{xy}$ is directly regulated by temperature. With increasing the value of *D*, the critical temperature corresponding to the sign reversal of $\kappa_{xy}$ increases. Notably, according to Eq. (17), the thermal Hall conductivity exhibits a dependence on the Berry curvature distribution and the occupation of the magnon bands. Thus, the discontinuous change in Berry curvature is directly reflected in the behavior of across $T_c$, where the discontinuous behavior is also reported in Ref. [34]. The discontinuity of on the temperature is noteworthy and experimentally diacritical. Experimentally, the magnon thermal Hall effect offers an alternative means of detecting the topological transition.

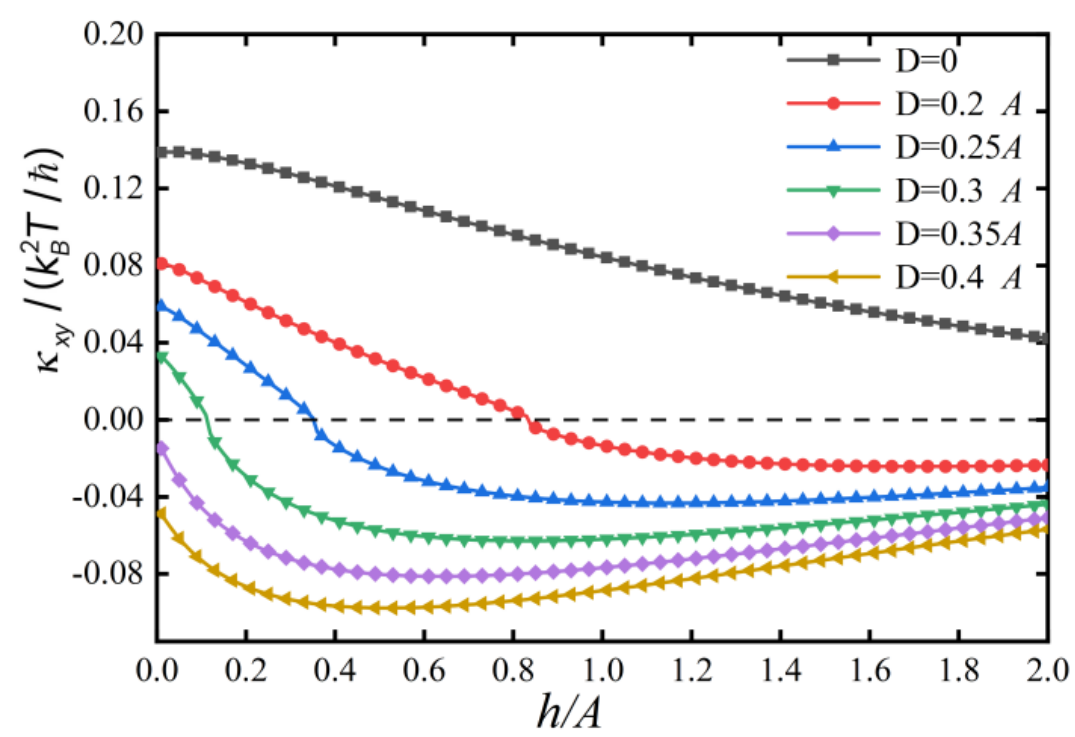

**Fig. 7.** The thermal Hall conductivity vs magnetic field strength for different values of *D*. The temperature is set to *T*=1*A*.

From the perspective of magnons, we also plot the thermal Hall conductivity as the function of the magnetic field across a spectrum of DMI strengths, as displayed in Fig. 7. For a fixed temperature, $\kappa_{xy}$ exhibits sign reversal with increasing magnetic field solely when the DMI is relatively weak. Importantly, as the value of D decreases, the critical magnetic field strength for the sign reversal of $\kappa_{xy}$ increases monotonically. Without the DMI, the magnon gap remains unclosed, as illustrated in Fig. 4(a). Consequently, it is not hard to understand that the sign reversal of $\kappa_{xy}$ cannot be attained in HK honeycomb ferromagnets lacking the DMI. Hence, the DMI plays a pivotal role in inducing the topological phase transition via temperature or magnetic field variation in HK honeycomb ferromagnets.

## 4. Conclusions

In summary, we have investigated the topological properties of HK honeycomb ferromagnets with DMI under the external magnetic field. The magnetic system exhibits distinct topological phase transitions driven by temperature or external magnetic field. We further present a series of rich magnon topological phase diagrams parameterized by DMI strength, temperature and magnetic field strength. Our results have shown that the introduction of DMI is a prerequisite for the model constructed in this study. Furthermore, the critical temperature of temperature-induced topological phase transitions monotonically approaches the Curie temperature with increasing DMI strength. These continuous topological phase transitions are driven by magnon-magnon interactions and are characterized by gap closing. The sign of the thermal Hall conductivity corresponds to the Chern number (a topological invariant), and its sign reversal during phase transitions serves as a reliable signature for identifying topological phases in subsequent experiments.

## Acknowledgments

This work was supported by the National Natural Science Foundation of China under Grant Nos. 12464013 and 12064011 and the Scientific Research Fund of Hunan Provincial Education Department under Grant No. 23A0404.

## References

[1] N. P. Armitage, E. J. Mele, and A. Vishwanath, Weyl semimetals as a platform for topological quantum matter, Rev. Mod. Phys. 90, 015001 (2018).

[2] B. A. Bernevig, T. L. Hughes, and S.-C. Zhang, Quantum spin Hall effect and topological phase transition in HgTe quantum wells, Science 314, 1757 (2006).

[3] I. Garate and M. Franz, Topological surface states in the presence of magnetic impurities, Phys. Rev. Lett. 110, 046402 (2013).

[4] L. Lu, J. D. Joannopoulos, and M. Soljačić, Topological photonics, Nat. Photonics 8, 821 (2014).

[5] M. S. Miao, Q. Yan, C. G. Van de Walle, W.-K. Lou, L. L. Li, and K. Chang, Topological surface states in half-Heusler compounds, Phys. Rev. Lett. 109, 186803 (2012).

[6] Y.-H. Li and R. Cheng, Thermal Hall effect of magnons in collinear antiferromagnetic insulators, Phys. Rev. B 103, 014407 (2021).

[7] R. Matsumoto and S. Murakami, Rotational motion of magnons and the thermal Hall effect, Phys. Rev. Lett. 106, 197202 (2011).

[8] I. Dzyaloshinsky, A thermodynamic theory of “weak” ferromagnetism of antiferromagnetics, J. Phys. Chem. Solids 4, 241 (1958).
[9] R. Jaeschke-Ubiergo, *et al.*, Magnon thermal Hall effect in kagome antiferromagnets with Dzyaloshinskii-Moriya interactions, Phys. Rev. B 103, 174410 (2021).
[10] S. A. Owerre, A first theoretical realization of honeycomb topological magnon insulator, J. Phys.: Condens. Matter 28, 386001 (2016).
[11] L. Zhang, J. Ren, J.-S. Wang, and B. Li, Topological magnon insulator in insulating ferromagnet, Phys. Rev. B 87, 144101 (2013).
[12] K. Li, C. Li, J. Hu, and Y. Li, Observation of the thermal Hall effect in a magnetic insulator, Phys. Rev. Lett. 119, 247202 (2017).
[13] K.-K. Li and J.-P. Hu, Thermal Hall effect of magnons in magnets with dipolar interaction, Chin. Phys. Lett. 34, 077501 (2017).
[14] P. A. McClarty, *et al.*, Topological magnons in pyrochlore ferromagnets, Phys. Rev. B 98, 060404 (2018).
[15] Y. H. Gao, *et al.*, Topological magnon bands and thermal Hall effect in frustrated magnets, Phys. Rev. Research 1, 013014 (2019).
[16] L.-C. Zhang, *et al.*, Topological magnon insulator in a two-dimensional ferromagnet, Phys. Rev. B 103, 134414 (2021).
[17] V. Brehm, P. Sobieszczyk, and A. Qaiumzadeh, arXiv:2409.15964 (2024).
[18] D. G. Joshi, Topological magnon bands in a tetragonal lattice, Phys. Rev. B 98, 060405 (2018).
[19] M. Deb and A. K. Ghosh, Topological magnon insulator in a one-dimensional ferrimagnetic chain, J. Phys.: Condens. Matter 31, 345601 (2019).
[20] S. S. Pershoguba, *et al.*, Magnon dark modes and gapless edge states in synthetic antiferromagnets, Phys. Rev. X 8, 011010 (2018).
[21] Y.-S. Lu, J.-L. Li, and C.-T. Wu, Topological magnon bands in a kagome lattice ferromagnet, Phys. Rev. Lett. 127, 217202 (2021).
[22] H. Zhu, H.-C. Shi, Z.-G. Tang, and B. Tang, Thermal transport properties of magnons in magnetic materials, Eur. Phys. J. Plus 138, 11 (2023).
[23] H.-C. Shi, *et al.*, Topological magnon transport in two-dimensional magnets, Phys. Lett. A 528, 130054 (2024).
[24] Y.-H. Li and R. Cheng, Thermal Hall effect of magnons in collinear antiferromagnetic insulators, Phys. Rev. B 103, 014407 (2021).
[25] L. Janssen, *et al.*, Magnon thermal Hall effect in noncollinear antiferromagnets, Phys. Rev. Lett. 117, 277202 (2016).
[26] T. Holstein and H. Primakoff, Field dependence of the intrinsic domain magnetization of a ferromagnet, Phys. Rev. 58, 1098 (1940).
[27] H. Katsura, N. Nagaosa, and P. A. Lee, Theory of the thermal Hall effect in quantum magnets, Phys. Rev. Lett. 104, 066403 (2010).
[28] R. Matsumoto and S. Murakami, Rotational motion of magnons and the thermal Hall effect, Phys. Rev. Lett. 106, 197202 (2011).
[29] T. Holstein and H. Primakoff, Field dependence of the intrinsic domain magnetization of a ferromagnet, Phys. Rev. 58, 1098 (1940).
[30] M. Bloch, Sur l'interaction entre les spins dans les matériaux ferromagnétiques, Phys. Rev. Lett. 9, 286 (1962).

[31] P.-O. Löwdin, A nonlinear orthogonalization procedure in general quantum mechanics, Phys. Rev. 97, 1474 (1955).

[32] K. Yosida, Theory of Magnetism, Springer (1996).

[33] Z. Li, T. Cao, and S. G. Louie, Topological magnon insulators in two-dimensional van der Waals ferromagnets, J. Magn. Magn. Mater. 463, 28 (2018).

[34] V. V. Mkhitaryan and L. Ke, Topological magnon bands in a square lattice antiferromagnet, Phys. Rev. B 104, 064435 (2021).

[35] R. Shindou, *et al*., Quantized thermal Hall effect of magnons in a pyrochlore ferromagnet, Phys. Rev. B 87, 174427 (2013).

[36] R. J. Birgeneau, *et al*., Spin waves in antiferromagnetic MnF2, Phys. Rev. B 5, 2607 (1972).

[37] O. W. Dietrich, *et al*., Spin dynamics in ferromagnetic EuO, Phys. Rev. B 14, 4923 (1976).

[38] R. Matsumoto, *et al*., Topological properties of magnon bands in a kagome lattice, Phys. Rev. B 89, 054420 (2014).

[39] S. M. Rezende, *et al*., Introduction to magnonics, J. Appl. Phys. 126, 151101 (2019).

[40] B. Wei, J.-J. Zhu, Y. Song, and K. Chang, Thermal Hall effect of magnons in a two-dimensional magnet, Phys. Rev. B 104, 174436 (2021).

[41] T. Hicks, T. Keller, and A. Wildes, Magnetic structure and spin dynamics of layered antiferromagnets, J. Magn. Magn. Mater. 474, 512 (2019).

[42] T. Fukui, Y. Hatsugai, and H. Suzuki, Chern numbers in discretized Brillouin zone: Efficient method of computing (spin) Hall conductances, J. Phys. Soc. Jpn. 74, 1674 (2005).